\documentclass[preprint]{emulateapj}

\slugcomment{Submitted to the Astrophysical Journal}
\shortauthors{Zaritsky, Zabludoff, \& Gonzalez}
\shorttitle{Equations of Galactic Structure}

\begin{document}
\title{Toward Equations of Galactic Structure}
  
\author{Dennis Zaritsky\altaffilmark{1}, Ann I. Zabludoff\altaffilmark{1}, and Anthony H. Gonzalez\altaffilmark{2}}
\altaffiltext{1}{Steward Observatory, University of Arizona, 933 North Cherry Avenue, Tucson, AZ 85721}
\altaffiltext{2}{Department of Astronomy, University of Florida, Gainesville, FL 32611}

\email{dzaritsky, azabludoff@as.arizona.edu, anthony@astro.ufl.edu}

\begin{abstract}    
We find that all classes of galaxies, ranging from disks to spheroids and from dwarf spheroidals 
to brightest cluster galaxies, lie on a two dimensional surface within the space
defined by the logarithms of the 
half-light radius, $r_e$, mean surface brightness within $r_e$, $I_e$, and
internal velocity, $V^2 \equiv ({1\over 2}v_c^2 + \sigma^2)$, where $v_c$ is the rotational velocity and
$\sigma$ is the velocity dispersion. If these quantities
are expressed in terms of kpc, $L_\odot$ pc$^{-2}$, and km s$^{-1}$, then this
surface is described by the equation $ \log r_e -  \log V^2 + \log I_e + \log \Upsilon_e + 0.8 = 0$,
where we provide a fitting function for 
$\Upsilon_e$, the mass-to-light ratio within $r_e$ in units of $M_\odot/L_\odot$, that depends only
on $V$ and $I_e$.
The scatter about this
surface for our heterogeneous sample of 1925 galaxies is small ($<$ 0.1 dex), and both
the scatter within one of the galaxy subsamples  (1319 disks) and
the analysis of subsamples with independently derived mass-to-light ratios suggest that
the intrinsic scatter could be as low as $\sim 0.05$ dex, or 10\%, prior to any correction for
observational errors.  This small scatter has three possible implications for how gross
galactic structure is affected by 
internal factors, such as stellar orbital structure, nuclear activity, or mass
loss history, and by external factors,
such as environment or accretion history. These factors either
1) play no role beyond generating some of the observed scatter, 2) move galaxies 
along the surface, or 3) balance each other to maintain this surface  
as the locus of galactic structure equilibria.
We cast the behavior of $\Upsilon_e$ in terms
of the fraction of baryons converted to stars, $\eta$, and the concentration
of those stars within the dark matter halo, $\xi \equiv R_{200}/r_e$, where $R_{200}$ is the standard
estimate of the virial radius. We derive expressions for $\eta$ and $\xi$, use an independent measurement of $\eta$ to evaluate leading constant terms, obtain
$\eta = 1.9 \times 10^{-5} (L/L^*)\Upsilon_* V^{-3}$ and $\xi = 1.4 V r_e^{-1}$, and relate these 
to each other via $\log \eta + \log \xi = -\log \Upsilon_e + \log \Upsilon_* + const$.
Finally, we present the distributions of $\eta$ and $\xi$ for the full range of galaxies and conclude that the high $\Upsilon_e$'s of dSphs are due to low $\eta$ rather
than any differences in $\xi$, that $\eta$ is similar for spheroids and disks of a
given $V$, and
that $\eta$ decreases with increasing $V$ for systems with $V > 30$ km sec$^{-1}$.
For
systems with internal velocities comparable to that of the Milky Way ($149 < V <  163$ km s$^{-1}$) , 
$\eta = 0.14 \pm 0.05$, and
$\xi$ is, on average, $\sim$ 5 times greater for spheroids than for disks.

\end{abstract}


\section{Introduction}
\label{sec:intro}

Are galaxies fundamentally a simple family of collapsed objects, whose gross structure is
describable by a few basic parameters, or are they highly complex 
systems whose structural properties are determined by a myriad of internal and external factors?

If the former, there must be an analogous 
construct to the
stellar Hertzprung-Russell diagram that testifies to deep, systemic structural patterns among
galaxies and 
serves as a guide to a simple, if not entirely complete, analytic description of galactic structure.
The study of stellar structure offers a beautiful example
of the power of reductionism in astrophysics.
By focusing on the HR diagram, investigators solved
the problem of stellar structure without needing to address other unsolved problems, such 
as the origin of the initial mass function.
The observation that the position of main sequence stars on the HR diagram is insensitive
to their location in the Galaxy indicates that
their structure does not depend sensitively on 
parameters that vary from one place to another.
We now know that mass is the primary determinant of where a star lies on the main sequence.
Other physical characteristics,
such as age, metallicity, and rotation, affect stellar colors and magnitudes 
(and therefore should be included in a complete model of stellar structure), but 
they are relatively minor factors along the main sequence 
and can be neglected in the interest of isolating the 
basic physics. 

Among galaxies there are hints of analogous ``sequences".
These are referred to as galaxy scaling laws and include
the Faber-Jackson \citep[FJ;][]{fj} and  
Tully-Fisher relations \citep[TF;][]{tf}, the Fundamental Plane \citep[FP;][]{dd87, d87}, 
and the Fundamental Manifold \citep[FM;][]{zgz,zgzb}.
Although it is not yet evident
that any of these is as fundamental for galaxies as the main sequence is for stars, 
they do imply that a limited number of parameters  characterize 
the gross properties of at least certain subsets of galaxies.

There are two arguments against using existing galaxy scaling laws as guides to a fuller
description of galactic structure. 
First, existing scaling laws work only
over a limited range of galaxy types and luminosities. While this is not 
an insurmountable obstacle --- not all stars lie on the main sequence --- it
suggests that the current scaling
laws are incomplete and that they will not lead to a 
description of all galaxies. 
Second, for historical reasons related to the empirical nature of the scaling laws,   
their current formulation 
is not optimal with respect to possible theoretical constructs. 
For example, one determination of the Fundamental Plane has
$\log r_e = 1.24 \log \sigma - 0.82 \log I_e + \gamma$ \citep{jorg}, where 
$r_e$ is the half-light radius, $\sigma$ is the velocity dispersion, $I_e$ is
the surface brightness within $r_e$, and $\gamma$ is a constant. 
It is unlikely that a simple theory 
would reproduce the 1.24 and 0.82 coefficients.

In this paper, we attempt to address both of these shortcomings. Although we are not the first
to hope to identify a unifying description of galaxies \cite[see $\kappa$-space;][]{burstein}, we 
achieve three of our key goals: 1) to find an empirical relationship for {\sl all} galaxies that
has comparable scatter to those relations identified previously for limited subsets of galaxies
(TF, FJ, FP, and FM), 2) to
isolate the critical additional knowledge beyond the virial theorem that is needed to derive
this relationship, and 3) to begin constructing the bridge between a purely empirical 
relationship that utilizes observables and a theoretical one that is based
on physical parameters.

In summary, we 
begin from basic dynamical principles and examine the dimensionality of the family of 
galaxies, ranging from dSph's to brightest cluster galaxies and from disks to spheroids.
We address our basic question --- how uniform are the gross structural properties of galaxies? ---
by determining that a single scaling relation exists that spans all luminosities
and galaxy types and by quantifying its scatter.
We employ an extended version of the fundamental plane formalism that 
reproduces the structural properties of all galaxies at a level comparable to
that achieved with either the TF relation for disks or the FP
for spheroids. We establish that connecting this new scaling relation to the virial theorem 
requires knowledge only of the mass-to-light ratio within $r_e$, $\Upsilon_e$, and that
$\Upsilon_e$ can be accurately modeled as a function of the observed structural parameters
themselves. We proceed to 
describe $\Upsilon_e$ as a combination of the fraction of baryons
converted into stars, $\eta$, and the degree to which those stars are packed within
the dark matter halo, $\xi$. Using our empirical findings, we then  calculate
these two physical parameters for all of our galaxies and compare $\eta$ with
independent measurements. The aim of this work is to define
simple expressions for basic physical parameters of galaxies  that may illustrate 
which physical processes 
drive the observed patterns of galactic structure, 
with the expectation that this will focus subsequent, more detailed theoretical
work.

\section{The Data}
\label{sec:data}

To determine the degree to which all galaxies are structurally similar, we need 
structural parameter measurements for galaxies ranging from spheroids to disks and giants to dwarfs.
Part of the legacy of distinct scaling relations for different classes of galaxies are 
studies that provide the relevant information only for those particular classes of galaxies. 
For example, there are extensive studies of 
spheroidal galaxies \citep[e.g.,][]{jorg} that are entirely
distinct from those of spirals \citep[e.g.,][]{springob}. This dichotomy is partly
due to the techniques necessary to measure the internal dynamics for disks and spheroids, 
but it also leads to the use of different photometric systems and definitions. It is impossible to resolve all of those differences, and many existing galaxy samples cannot be included here because they lack some
necessary measurements.
We describe the spheroid and disk samples that we use below. These constitute a heterogeneous dataset, 
but span the full range of galaxy types and luminosities, and require minimal corrections
for internal comparisons. It is a testament to the robustness of our results
that the many differences among the samples that we either ignore or
only crudely correct (such as correcting the photometry to $I$ band on 
the basis of average colors for different galaxy populations) 
do not derail this
investigation.

\subsection{Spheroids}
\label{sec:spheroids}

Since \cite{zgzb}, there has been one key improvement in the available
data on low-mass spheroids. \cite{simon} present velocity dispersions, and a uniform set of
structural parameters, for eight additional Local Group dSph's, including some of the lowest
luminosity systems known. Adding these data to the \cite{zgz} compilation 
greatly increases our sample for extreme
values of  luminosity, internal velocity, and effective radius.
The lack of such data earlier precluded our use of this range
in the {\sl fitting} of the FM, and instead we 
showed that an extrapolation of the FM accurately fit galaxies in this parameter range  \citep{zgzb}.
Here, we fit to both the previous data for the entire range of spheroid masses
\citep[][and references therein]{zgz, zgzb} and the new data for low mass spheroids \citep{simon}.

\subsection{Disks}
\label{sec:disks}

We focus on three particular disk
samples: \cite{pizagno}, \cite{springob}, and \cite{geha2006}.
Here we briefly describe the various data sets.

Of the three samples, the \cite{pizagno} sample allows the simplest comparison to the spheroid
samples.
The authors provide half-light radii, $i-$band magnitudes, and a range of velocity 
measures from their optical rotation curves. 
As they did for their Tully-Fisher analysis, we use their $V_{80}$ measurement,
which is a measure of the rotation velocity at a radius that encloses 80\% of the galaxy light.
We correct the $i-$band magnitudes to Johnson by subtracting 0.4 mag \citep{fukugita}.

The next simplest sample for comparison is that of \cite{springob}, who provide
HI measurements of the rotation and $I$-band photometry. They do not tabulate half-light
radii, so we calculate them based on the measures they do provide, the radius
that encloses 83\% of the light and the radius of the 23.5 mag (sq. arcsec)$^{-1}$ isophote,
assuming an exponential surface brightness profile. Among galaxies for which
all of the relevant data exist, we only reject
systems with $cz < 2500$ km sec$^{-1}$, to avoid the local flow field.

Lastly, the \cite{geha2006} sample is distinct because it
is primarily composed of low luminosity systems with very large gas mass fractions. 
Because gas fractions are low in normal spirals \citep{read}, the
gas can be ignored with little impact when studying scaling relations like TF for such spirals. 
However, studies of low mass galaxies show that accounting for all the
baryons is critical in maintaining the scaling relation \citep{mcgaugh00,mcgaugh05,geha2006}.
Therefore, we discuss the \cite{geha2006} sample separately in \S \ref{sec:evolving}.
We use their inclination- and turbulence-corrected velocities, and transform from $r$ to $I$ magnitudes
using the colors of late type spirals and the tabulations of \cite{fukugita}.

\section{Results and Discussion}
\label{sec:discussion}
\subsection{Proceeding from the Virial Theorem}
\label{sec:beyond}

In this section, we revisit the standard derivation of the FP 
to provide a framework and physical intuition for our observational results.
We begin with the tensor virial theorem, use simplifying assumptions to rewrite
the virial theorem in terms of observed quantities where possible, discuss the
resulting equation and its
implications for the nature of galactic structure, and finally
suggest a way to proceed even though some terms in the resulting equation cannot be expressed
in terms of observed quantities. We apply this suggestion and explore it in quantitative detail in \S \ref{sec:simplicity} and \S \ref{sec:validity}.

We begin with the tensor virial theorem, which is
\begin{equation}\label{eq:1}
{1\over 2}{d^2I_{jk}\over dt^2} = 2T_{jk} + \Pi_{jk} + W_{jk},
\end{equation}
where
$\bf{T}$ and ${\bf \Pi}$ are the contributions to the kinetic energy tensor from the ordered 
and random motions, ${\bf W}$ is the potential energy tensor, and ${\bf I}$ is the moment of inertia
tensor.  In steady 
state, the left hand side of Eq. \ref{eq:1} is zero. We evaluate the trace of this equation and
express the ordered component of the kinetic energy as ${1 \over 2}A_0 M v_c^2$ and
the random as $A_1 M\sigma^2$,  where $v_c$ is the circular velocity in disk galaxies,
$\sigma$ is the line-of-sight velocity dispersion for spheroidals, $M$ is the mass of the system, 
and the $A$'s represent the correction factors obtained by fully evaluating the appropriate integrals.
Similarly, the potential energy is expressed as $-B_0GM^2/R$, where
$R$ is a characteristic radius that we define to be the half-light radius, $r_e$, and
$B_0$ is a correction factor obtained by fully evaluating the appropriate integral.
Hence, without loss of generality, 
\begin{equation}\label{eq:2}
A_0v_c^2 + A_1\sigma^2 = B_0{GM \over r_e}.
\end{equation}
All of the possible real-world complications are encapsulated in the yet unspecified $A$'s and in $B_0$,
and, in principle, these could be extremely complicated functions of the formation history and
environment of  galaxies. The only assumption that we have made so far is that the virial theorem
holds over these radii, which is reasonable for galaxies because $r_e \ll r_{vir}$.

To numerically evaluate Eq. \ref{eq:2}, 
we now introduce two sets of simplifying assumptions that we will eventually 
test by determining whether we reproduce the observations.
First, is the  {\sl kinematic} simplification. We reduce the number of $A$ parameters, by requiring 
$A_0v_c^2 + A_1\sigma^2 \equiv AV^2$, where $V \equiv ({1 \over 2} v_c^2 + \sigma^2)$. This simplification
is accurate if we are dealing with isothermal spheres and isotropic velocity dispersions. In such
systems, at large radii, $v_c$ for a purely rotationally-supported population equals $\sqrt{2}\sigma$
for a purely pressure-supported population.
Furthermore, in such a pressure-supported
system, the internal velocity dispersion is equal to the line-of-sight velocity dispersion. 
To evaluate $V$ for disk galaxies, we will use the measured $v_c$ and set $\sigma=0$, while for spheroidal galaxies we will use the measured $\sigma$ and set $v_c = 0$. 

Although the kinematic simplification 
relies on highly specific assumptions, both disks and spheroids satisfy the relevant
conditions well, and this conversion has been used previously 
in various contexts \citep{burstein,kassin}. Optical disks are characterized by flat rotation curves, which 
imply that the mass profile is that of an isothermal sphere over these radii,
and that the velocity tracers, H II regions or neutral hydrogen, are on circular orbits \citep[see][]{faber}.
Spheroids also lie in mass distributions that are consistent with being
isothermal spheres \citep{gavazzi}, and their stellar velocity dispersions are 
nearly isotropic if the system is a slow rotator \citep{cappellari2007}. Due to
the nature of our spheroid samples, we expect that there are strong selection biases
against fast rotators among the more luminous systems (i.e., they would often be removed from 
Fundamental Plane studies), and the lowest luminosity systems show little rotation \citep{walker}.
Therefore, our sample is likely to satisfy the assumptions involved in the kinematic simplification, 
but we discuss possible signatures of failure in \S \ref{sec:revisiting}.

\begin{figure}
\plotone{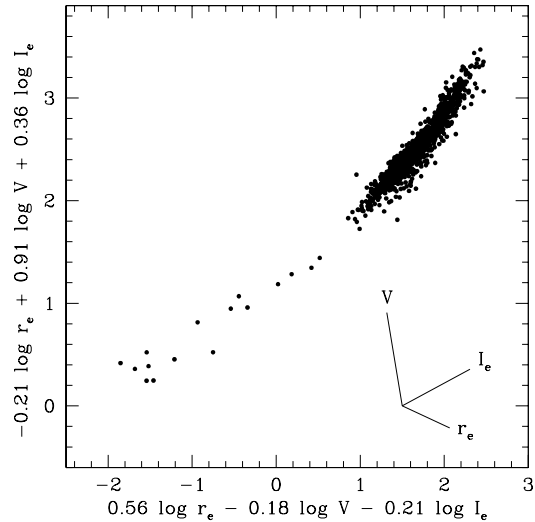}
\caption{The distribution of galaxies (444 spheroids, 1481 disks) in the ($r_e,I_e, V$) space.
This projection was chosen by eye to demonstrate that galaxies lie on a fairly 
well defined surface. The data are discussed in \S \ref{sec:data}.
The axes in the lower right show the orientation of the 3-space.
The poorly populated tail consists of Local Group dwarfs and would be, we expect, well
populated in a volume limited sample. Units for the three axes are kpc, $L_\odot$/pc$^2$, and
km sec$^{-1}$, where $V={v_c \over \sqrt{2}}$ for disks and $V=\sigma$ for spheroids.}
\label{fig:all}
\end{figure}

Continuing in our attempt to convert Eq. \ref{eq:2} into an equation that we can numerically
evaluate, our
second simplification involves the 
replacement of the difficult-to-measure $M$ with $M_e$, the mass enclosed 
at $r_e$. We refer to this as the {\sl mass} simplification.
We rewrite $M_e$ as $\Upsilon_e L_e$ so that it is expressed as a function of
observable quantities: 
$L_e$ and $\Upsilon_e$, the luminosity ($\equiv \pi r_e^2I_e$) and the mass-to-light ratio 
within $r_e$, respectively. We then
replace $B_0$ with $B$ to account for the unknown difference between $M$ and $M_e$.

We now apply the kinematic and mass simplifications to Eq. \ref{eq:2}, rewriting it as 
\begin{equation}\label{eq:3}
AV^2 = B{G\pi\Upsilon_e r_e I_e}.
\end{equation}
 Finally, we take the logarithm of both sides 
and rearrange terms to obtain
\begin{equation}\label{eq:4}
\log r_e -  \log V^2 + \log I_e + \log \Upsilon_e - \log A +  \log B = const.
\end{equation}

This equation leads to the rather dispiriting conclusion that galaxies populate at
least a six dimensional parameter space --- more if yet unspecified parameters, such as $\Upsilon_e$,
are actually functions of additional 
parameters, like age, metallicity, formation history, bulge-to-disk
ratio, or environment.
Surprisingly, as shown in Figure \ref{fig:all}, galaxies  
populate a limited region of the $(r_e,I_e,V)$ space, indicating a much lower dimensionality.

One way in which the dimensionality of the galaxy family might be reduced from that
suggested by Eq. \ref{eq:4} 
is if $\Upsilon_e$, $A$, and $B$ are functions {\sl only} of $r_e, I_e$, and $V$.
A simple variant of this scaling is referred to as ``homology", in which the functional
forms are assumed to be power laws. Because of the logarithms, the end effect of rewriting
Eq. \ref{eq:4} in such a variant is
a change in the coefficients of the $\log V^2$ and $\log I_e$ terms.
Therefore, the assumption of homology results in a prediction that galaxies lie on a plane in the 
$(r_e,I_e,V)$ space. The values of the 
coefficients describe the tilt of that plane.
The success of the Fundamental Plane
description for giant ellipticals \citep{dd87,d87,jorg, bernardi} demonstrates that, over the limited
mass range of these galaxies, the homology assumption holds surprisingly well. This success
was extended 
in the $\kappa-$space formalism of \cite{burstein}, where different classes of objects
were found to lie on different planes. However, the failure of a single
plane to describe the distribution of all 
spheroidal galaxies demonstrates that over
a more extended mass range,  which includes the most and least massive spheroids, 
homology does not hold \citep{zgz}.

Here we take a different approach in that we (1) assume that {\sl all} galaxies (faint, luminous,
disk, and spheroid) fall on a single manifold in the $(r_e,I_e,V)$ space and (2) examine
the behavior of $\Upsilon_e$ that would make that possible. 
This approach is motivated by our earlier finding that the behavior of $\Upsilon_e$ is both
simple and qualitatively reasonable for all spheroids \citep{zgz,zgzb} --- leading to what we 
termed the Fundamental Manifold (FM) of spheroids.
So emboldened, we now assert that for all galaxies deviations from homology are
dominated by the behavior of  $\Upsilon_e$, 
ignoring variations in $A$ and $B$ among galaxies\footnote{An alternate treatment
that eventually leads to the same conclusion is to group together $\Upsilon_e$, $A$, and $B$ into
a generic unknown, $\Delta$, fit for $\Delta$, and then use the argument in \S \ref{sec:validity} to demonstrate
that $\Delta \propto \Upsilon_e$.}. 
Our approach here represents a philosophical departure from ours and others' earlier work, 
which usually focused on establishing or quantifying the tight empirically-derived
scaling relationships (e.g., the FP or FM), because we posit the
existence of a fundamental manifold of all galaxies and then examine the implications.

\begin{figure*}
\plotone{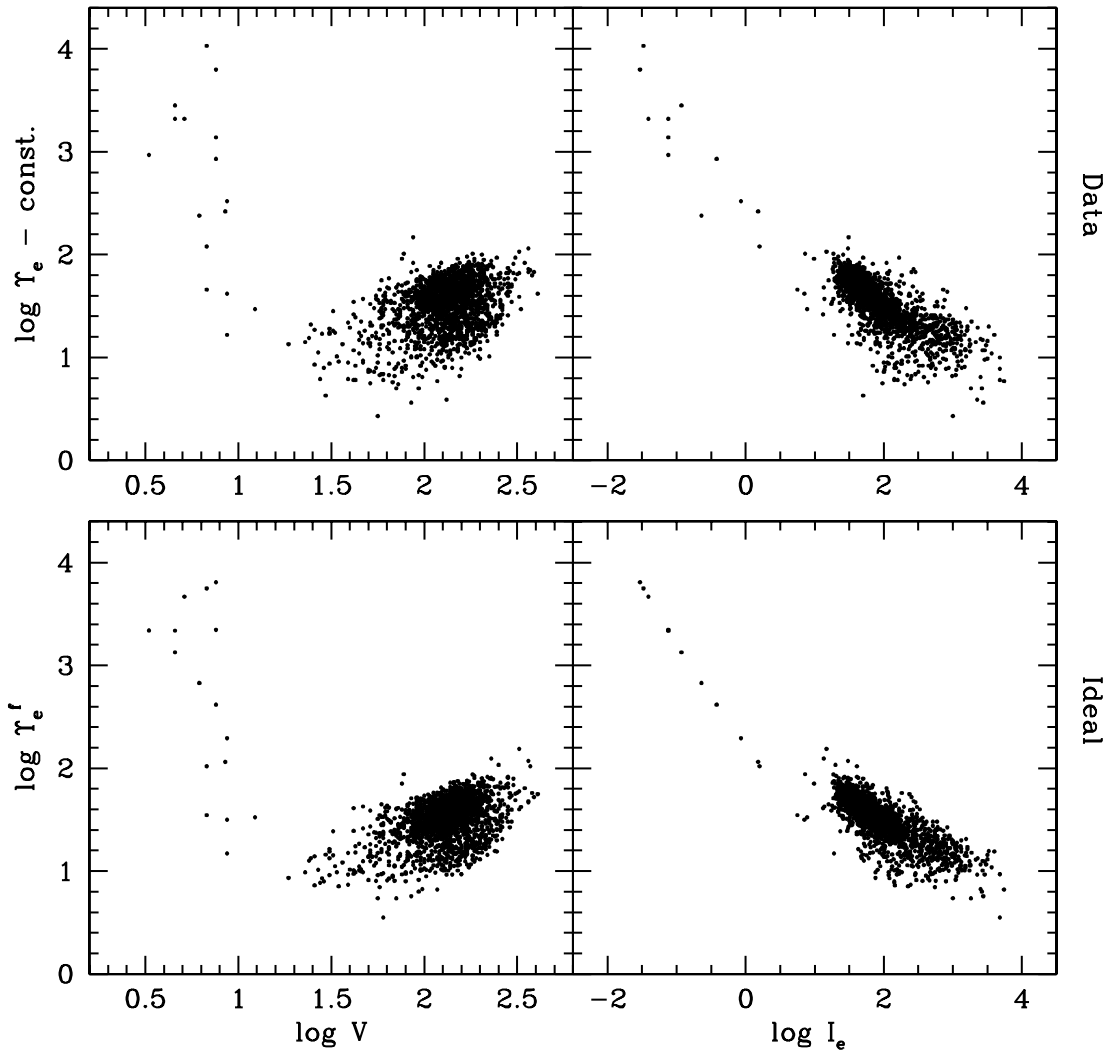}
\caption{Projections of $\log \Upsilon_e$ and $\log \Upsilon_e^f$. We plot the projections of $\log \Upsilon_e - const.$,
determined from Eq. \ref{eq:5},  vs. $\log V$ and $\log I_e$ in the upper panels. 
In the lower panels, we plot the values of $\log \Upsilon_e^f$ for every galaxy in our sample
using the fit given in Table 1. The lower
panels illustrate how even with no intrinsic scatter in $\log \Upsilon_e^f$ the projections
show significant apparent scatter. We conclude that the bulk of the apparent
scatter in the upper panels is due to the effects of projecting the complicated surface onto these
axes rather than observational errors or intrinsic scatter. }
\label{fig:projections}
\end{figure*}

\begin{figure*}
\plotone{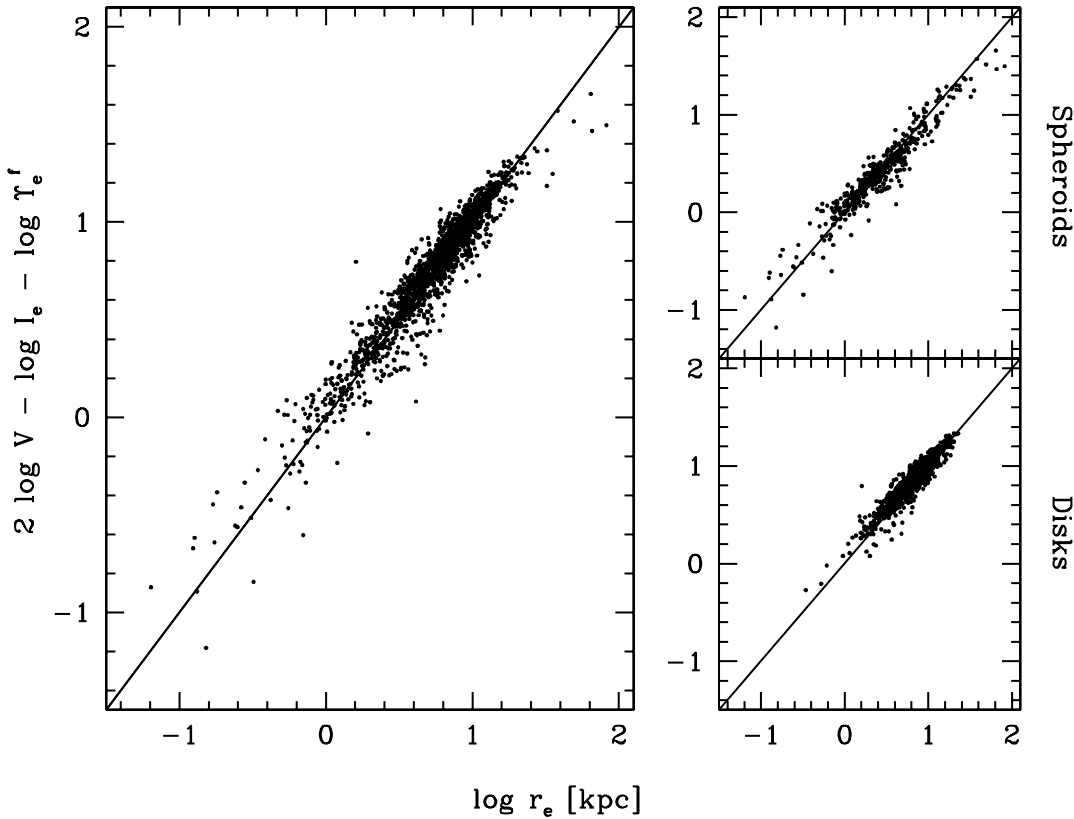}
\caption{Testing an equation of galactic structure, Eq. \ref{eq:5}. We replace $\log \Upsilon_e - const.$ 
with $\Upsilon_e^f$ to evaluate Eq. \ref{eq:5} for the entire sample (left panel) and 
for spheroids and disks separately (right panels).   By construction, the mean relation should 
lie along the 1:1 line. The low scatter and the lack of systematic deviations for galaxy subsamples
testifies to the universal nature of this simple relation.
}
\label{fig:fm}
\end{figure*}

\subsection{The Simplicity of Galaxies}
\label{sec:simplicity}

The treatment described in \S \ref{sec:beyond} and culminating in Eq. \ref{eq:4} is 
incomplete. The simple
theoretical approach fails because it does not predict the low dimensionality of the data seen
in Figure \ref{fig:all}. On the other hand, the purely empirical treatment
of fitting a manifold to
the data in the $(r_e,I_e,V)$ space fails because it 
does not connect the actual functional form to a physical framework. Much like the case with
the FP coefficients, directly fitting the data will
subsume the behavior of $\Upsilon_e$, $A$, and $B$ in Eq. \ref{eq:4} into the
coefficients of the various structural terms (see axes in Figure \ref{fig:all}).
Instead, we merge the two approaches by retaining 
the values of the coefficients derived from the virial theorem treatment as given
in Eq. \ref{eq:4}, assert that the most distinct break from homology occurs in $\Upsilon_e$, 
set $A$ and $B$ to be constants, and then solve for $\Upsilon_e$, 
\begin{equation}\label{eq:5}
\log \Upsilon_e = \log V^2 - \log I_e - \log r_e  + const.,
\end{equation}
This approach may seem like only mathematical sleight-of-hand, but we will quantitatively
test our association of $\Upsilon_e$ with the dominant departures from homology in \S \ref{sec:validity}.

To proceed, we
evaluate $\log \Upsilon_e -\ const.$ using Eq. \ref{eq:5} and plot the results in 
Figure \ref{fig:projections}. We then fit for the function, $\log \Upsilon_e^f$, that describes
these data and also plot the calculated values using this fit in Figure \ref{fig:projections}. Because
we are fitting to a distribution of points in a 3-space, 
the fitting function will depend on two variables, and the natural choices are those
that are distance independent, $V$ and $I_e$.
We want to minimize 
the fitting order, while still capturing the 
behavior
of the distribution. As demonstrated by \cite{zgz}, there is at least a second-order
dependence on $\log \sigma$ and some dependence on $\log I_e$, 
and so we fit to second-order in both $\log V$
and $\log I_e$ and include cross-terms.  
For this fit, we use only a
randomly selected one-sixth of the \cite{springob} sample to avoid having that sample dominate
the fit. 
We present the coefficients of our
fit in Table 1, but, because of the heterogeneous nature of the data and our
avoidance of any type of Malmquist-like corrections \citep{willick}, these
numbers are far from definitive. 

The distinction between this work, with its complex characterization of $\Upsilon_e$, 
and either FP or $\kappa$-space, with their assumption of homology, becomes evident
when examining Figure \ref{fig:projections}.
The projections of the data and the fitting function in Figure \ref{fig:projections}
illustrate how even in projection the functional form that describes $\Upsilon_e$ deviates
from power laws. The upper
panels contain the inferred values of $\log \Upsilon_e - const.$ from Eq. \ref{eq:5} vs. either $\log V$ or $\log I_e$. 
The lower panels, which show the fitted values, $\log \Upsilon_e^f$, and therefore have
no intrinsic scatter, illustrate how the bulk of the observed
scatter in the upper panels comes simply from the projection of a complicated surface
onto these axes. In other words, the reason why galaxies of the same $V$ have a range
of $\Upsilon_e$'s is not primarily because there is intrinsic scatter --- say, due to age or metallicity --- but
rather because galaxies have a range of $I_e$. For a given $V$ and $I_e$, the scatter
in $\Upsilon_e$ is much smaller than that observed in Figure \ref{fig:projections}. To be
specific, the scatter for the entire sample about the fit is 0.094 dex (24\% rms in $\Upsilon_e$). In 
contrast, the observed scatter in $\log \Upsilon_e$ in the upper left panel of Figure \ref{fig:projections} for
$1.9 < V < 2$ is 0.22 dex (66\% rms in $\Upsilon_e$).

\begin{deluxetable}{rrrrrr}
\tabletypesize{\scriptsize}
\tablecaption{$\Upsilon_e^f$ Fit Coefficients}
\tablewidth{0pt}
\tablehead{
\colhead{Constant} &
\colhead{$\log V$} &
\colhead{$\log I_e$} &
\colhead{$\log^2 V$ } &
\colhead{$\log^2 I_e$} &
\colhead{$\log V \log I_e$} \\
}
\startdata
2.12&$-$0.01&$-$1.05&0.07&0.13&0.14\\
\enddata
\label{tab:ml}
\end{deluxetable}

We are now ready to evaluate the degree to which Eq. \ref{eq:5} describes our set of galaxies.
Replacing $\log \Upsilon_e - const$ with $\log \Upsilon_e^f$, we evaluate Eq. 5 and plot
a rearrangement of the terms  in Figure \ref{fig:fm}.  By construction, 
Eq. \ref{eq:5} is satisfied on average when $\log \Upsilon_e - const.$ is replaced by $\log \Upsilon_e^f$, 
which is evident in Figure \ref{fig:fm}. 
The actual test of our approach comes from examining the scatter about the mean and whether
distinct galaxy populations fall off the mean trend. If
galactic structure depends strongly on parameters not included in this simple description, 
then the scatter will be large. In other words, two galaxies that are identical in the 
quantities $V$, $r_e$, and $I_e$ could, in principle, have very different values of $\Upsilon_e$ due
to a dependence of $\Upsilon_e$ on accretion history, age, varying degrees of mass loss, or
many other possible physical effects. These differences in $\Upsilon_e$ are not 
accounted for in $\Upsilon_e^f$, thereby potentially leading to a large scatter about the mean. 
However, the scatter is only 0.094 for the entire sample.
For reference, the scatter in this new relation for {\sl all} galaxies
is comparable to the scatter observed in either FP or TF studies for the relevant subset of galaxies.
The scatter can be reduced slightly, to 0.084, if we 
correct each galaxy sample separately for zero point shifts. These inferred zero
point shifts, obtained by calculating the mean offset relative to the 1:1 line, are all
comparable to plausible photometric errors, and generally correspond to 
a few hundredths of a magnitude. 

The success of placing all galaxies onto a single surface in the $(r_e,I_e, V)$ space
demonstrates that, to within the scatter ($< 25\%$), the family of galaxies is a two
parameter sequence, i.e., measuring two of these structural parameters specifies
the third. This implies that  potentially important factors in galaxy
development, such as environment, nuclear activity, star formation history, and 
accretion history, do not
play a significant role in determining galactic structure {\sl unless} they either move
galaxies along the surface in $(r_e,I_e,V)$ space or act in concert to preserve the
manifold as the locus of equilibrium points of galactic structure. 
To reiterate, the important aspect of Figure 3 is
not its linear nature, which is a result of our assertion that a fundamental manifold
exists for all galaxies, but rather the low scatter and lack of any systematic departures
for specific classes of galaxies, both of which imply that the assumptions that
we have made so far are appropriate to this level of precision. In other words, random or
systematic variations 
of $A$ and $B$ across either the mass range of galaxies or galaxy types, 
variations from isothermality, and any other
factors that are not considered contribute scatter that is at most the
observed scatter, which is comparable to that measured in TF or FP studies. 
We have achieved the first of the goals described in \S \ref{sec:intro} and now proceed
to examine whether our attribution of the departures from homology to $\Upsilon_e$ is
correct.

\begin{figure}
\plotone{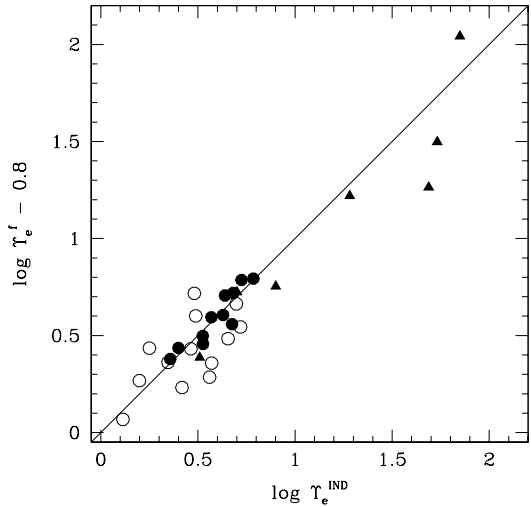}
\caption{A comparison of mass-to-light ratios derived from independent means, $\Upsilon_e^{IND}$ 
and our estimates of $\Upsilon_e$ using $\Upsilon_e^f$.  For normal ellipticals (circles), we compare
the mass-to-light ratio derived using the Schwarschild method of dynamical 
modeling for a set of nearby spheroidal galaxies \citep{cappellari} to our estimates
of $\Upsilon_e$ using $\Upsilon_e^f$. 
Open circles represent galaxies that \cite{cappellari} note are fast rotators and
filled circles represent those that are not. The line is the 1:1 correspondence. 
The filled circles, which are the most appropriate comparison sample, show only 0.06 dex
scatter (15\% in $\Upsilon_e$). For the Milky Way dSph galaxies (triangles), we compare the mass-to-light ratio derived from 
fitting NFW profiles to kinematic data \citep{walker} with our estimates
of $\Upsilon_e$ using $\Upsilon_e^f$. 
The unknown constant relating $\Upsilon_e^f$ to $\Upsilon_e$ is set to ensure
agreement in the mean values of $\Upsilon_e^{IND}$ and $\Upsilon_e^f$ and that value ($-$0.8) is then
adopted for Eq. \ref{eq:6}.
}
\label{fig:ml2}
\end{figure}

\subsection{The Physical Validity of  $\Upsilon_e^f$}
\label{sec:validity}

The mathematical trick of placing all of the galaxy formation physics beyond the virial theorem 
into 
$\Upsilon_e$ potentially masks the importance of $A$ and $B$. 
To check whether $\Upsilon_e^f$ truly reflects $\Upsilon_e$, or whether it
is in actuality a  composite of various terms,  we compare 
$\Upsilon_e^f$ to 
independent determinations of $\Upsilon_e$. We do this for both normal ellipticals and
dSphs. 

First,
we compare to values of $\Upsilon_e$ derived from a full Schwarschild 
analysis of the 2D line-of-sight velocity
distributions, $\Upsilon_e^{Sch}$, of normal ellipticals \citep{cappellari, cappellari2007}. 
Because of the unknown constant in the definition of $\Upsilon_e^f$ relative to $\Upsilon_e$,
we have 
the freedom to normalize $\Upsilon_e^f$ to best match the \cite{cappellari} data, which we do below (Figure 
\ref{fig:ml2}). 
Figure \ref{fig:ml2} illustrates the excellent correspondence between 
$\Upsilon_e^f$ and $\Upsilon_e^{Sch}$. 
The agreement is particularly good (0.06 dex rms, 15\% in $\Upsilon_e$)
for the galaxies that are most appropriate for our construction, namely those
where the velocity dispersion dominates over systemic rotation and
anisotropy measures are small ($|\beta, \gamma, \delta| < 0.15$ as 
measured by \cite{cappellari2007}). The
scatter for those with large anisotropies is significantly greater (0.17 dex rms, 48\% in $\Upsilon_e$),
suggesting that a full knowledge of $A$ and $B$ would decrease the scatter in Figure \ref{fig:fm} 
among those galaxies that do not fully satisfy the basic assumptions of our approach. 

\begin{figure}
\plotone{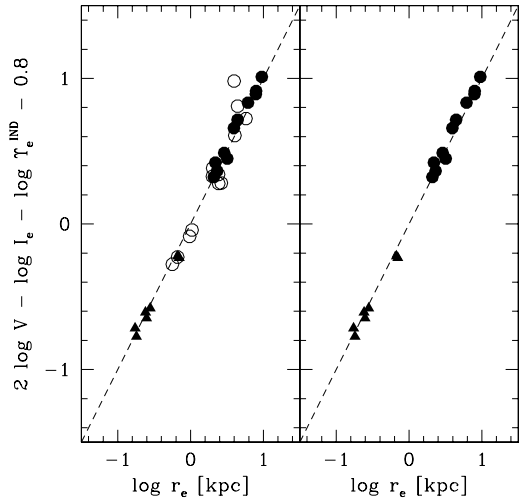}
\caption{The scaling relationship, Eq. \ref{eq:6}, using $\Upsilon_e^{IND}$ 
in place of $\Upsilon_e$.
Data and symbols are as in Figure \ref{fig:ml2}. The left panel includes all of the spheroids
with $\Upsilon_e^{IND}$, and the scatter  is 0.09 dex about the 1:1 line with a mean offset of 0.005.
In the right panel, we have removed the ellipticals with either significant rotation or
anisotropy \citep{cappellari,cappellari2007}, and the scatter drops to 
0.04 dex about the 1:1 line with a mean offset of $0.004$.}
\label{fig:mlfm}
\end{figure}

Second, we examine whether this correspondence holds across the range of galaxy masses.
In Figure \ref{fig:ml2} we include values of $\log \Upsilon_e$ for Galactic dSphs 
estimated using NFW model fits to the extensive kinematic data of \cite{walker}\footnote{The calculated $\Upsilon_e$'s use masses enclosed
within $r_e$ as calculated from the published fits, courtesy of Matthew Walker.}.
Unlike the Schwarschild 
analysis of normal ellipticals, 
which has the freedom to include anisotropic velocity distributions, the \cite{walker}
analysis does not. 
The dSph data suggest
a slightly different constant offset between $\log \Upsilon_e^f$ and $\log \Upsilon_e$ 
($-$0.78 rather than $-$0.82), but this
change is modest and has a nearly undetectable effect on the scaling
relation when ignored (see Figure \ref{fig:mlfm}).
We adopt the average offset, $-$0.8, as the normalization constant and suggest
an uncertainty in this number of $ \sim$ a few hundredths.
Using this correspondence,
we replace the constant in Eq. \ref{eq:5} to obtain
\begin{equation}
\label{eq:6}
\log r_e -  \log V^2 + \log I_e + \log \Upsilon_e + 0.8 = 0,
\end{equation}
where one can either evaluate $\Upsilon_e$ in some independent manner or
express it in terms of $V$ and $I_e$ using the fit given in Table 1 and replacing 
$\log \Upsilon_e + 0.8$ with $\log \Upsilon_e^f$.

The applicability of the same normalization
constant for both normal ellipticals and dSphs 
supports the contention that structural variations, as reflected by
changes in $A$ and $B$, are modest over most of the mass scale covered in
Figure \ref{fig:fm}. 
Using these independently derived measures of $\Upsilon_e$, we now return to Eq. \ref{eq:6},
use the literature values for $\Upsilon_e$ rather than our fitting function, and
plot the results in Figure \ref{fig:mlfm}. The difference between the left and right panels is 
the exclusion of the ellipticals that show evidence for rotation or anisotropic
velocity dispersions. The data in both panels follow the 1:1 correspondence well, although
the correspondence in the right panel is striking.
The scatter in that panel is 0.04 dex, or less than 10\% in the parameter values themselves. 
We conclude that 
to a level of precision between 10 and 25\% (the scatter measured using these independently
measured values of $\Upsilon_e$ and the scatter measured using our fitting function, $\Upsilon_e^f$,
respectively),
$\Upsilon_e$ encompasses
all of the additional physics necessary to proceed from the virial theorem to a description
of galactic structure. Thus, we achieve the second goal listed in \S \ref{sec:intro}.

\subsection{Evolving onto the Manifold}
\label{sec:evolving}

For various reasons, the small scatter seen in Figures \ref{fig:fm} and \ref{fig:mlfm} is remarkable. 
Even if $r_e$ and $V$ are the same in two similar galaxies, one
might expect variations in the stellar mass-to-light ratios, $\Upsilon_{*}$, of more than 50\%, which 
would introduce scatter via variations in $I_e$. 
We suspect that at least part of the reason for the low observed scatter lies in the selection
of galaxies in TF and FP studies, which generally favor  evolved, dynamically-relaxed galaxies, 
which are unlikely to exhibit dramatic variations in $\Upsilon_{*}$.  Therefore, we now return 
to the last of our disk galaxy samples \citep{geha2006}, which contains disks with
extremely high gas mass fractions, and might therefore be expected to harbor systems
with dramatically different values of $\Upsilon_{*}$ and $\Upsilon_e$ than those 
included in our analysis so far.

\begin{figure}
\plotone{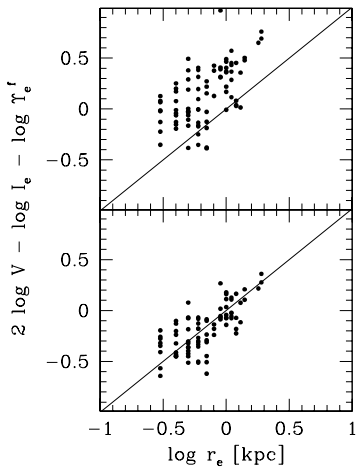}
\caption{Results for the \cite{geha2006} sample of disk galaxies with high gas mass fractions. 
Upper panel shows the galaxies  (points) in
comparison to the fitted relationship (line). Lower panel shows to where the galaxies
might evolve as their gas is turned into stars.}
\label{fig:geha}
\end{figure}

When the majority of the baryons in a galaxy are in the gas rather than in stars, 
one might expect that our treatment
as described above will fail because the connection between optical luminosity and mass
via $\Upsilon_e$ becomes tenuous. 
In fact, the galaxies in the \cite{geha2006} sample do fall off the surface, as shown 
in the upper panel of Figure \ref{fig:geha}. However, as demonstrated with regards
to the Tully-Fisher relation \citep{mcgaugh05, geha2006},  gas-rich and gas-poor 
galaxies have consistent scaling
relationships if one considers total baryonic content 
instead of just that in the stellar component. 
Reviewing Eqs. \ref{eq:2} and \ref{eq:3}, it is evident that the derivation of 
later equations, such as Eq. \ref{eq:6}, depended on a 
proxy for the enclosed mass. In light of  previous studies, while optical luminosity may be an
adequate proxy when most of the baryons are in stars, it is clearly inadequate 
when the majority of baryons are in the gas (or, more precisely, our fitting formula 
for $\Upsilon_e$ fails when different fractions of the baryons are luminous in otherwise similar
galaxies). 
The reason optical luminosity works well for most disks is
that the majority of their cold gas has been turned into stars \citep{roberts, bothun, kannappan, 
read}. Given that we do not have measured colors and gas masses for most of our sample, 
we cannot reformulate everything in terms of baryonic mass, but we can ask 
where these gas-rich galaxies would lie with respect to the 1:1 relationship
in Figure \ref{fig:geha} if they turned their gas into stars. Can they evolve onto the manifold?

To complete this exercise, we make two questionable assumptions. 
First, we assume that $r_e$ does not change during the conversion of gas to stars. 
Because the gaseous and stellar radial
distributions are likely to be different, this assumption must fail at some level.  Second, we 
assume that the stellar population formed from the gas eventually has the same 
stellar mass-to-light ratio (i.e., for stars only)
as the late-type galaxies in our sample
\citep[$\Upsilon_*= 0.97$ in the $I$-band for a typical Sbc galaxy with $B-V$ = 0.57;][]{fukugita,bell}.
For these two assumptions, we then find the fraction of gas turned into stars that
produces the best match to our scaling relation (Eq. \ref{eq:6}). 
We exclude one galaxy with a rotation velocity of 5 km sec$^{-1}$, which appears to be 
unphysically small, and one galaxy with $b/a > 0.9$, for which it is difficult to deproject
the velocity width.

The result of this exercise, that these high gas mass fraction galaxies can evolve onto the relationship
defined by the larger sample, is 
shown in the lower panel of Figure \ref{fig:geha}. The resulting scatter is 0.16 dex, which is
slightly lower than that found for the dSphs, but larger than for all of the galaxies
combined. The difficulty with our scenario is that
the agreement requires turning about  20\% more gas than is available throughout the galaxy
into stars. Perhaps this failure reflects the need for infalling gas, but it may also mean
that as the galaxy evolves there is a corresponding change in $r_e$. For example, in a model
where the gas is funneled efficiently into the center so that $r_e$ is only one-third
of its current value, and 75\% of the current gas is turned into stars, the scatter about
the relationship is only 0.14 dex. We conclude that there are plausible scenarios
in which these gas-rich galaxies lie on the observed relationship and that one might identify
galaxies that are still strongly evolving or those with a significant reservoir of cold gas as
outliers from the manifold. Nevertheless, if these galaxies either convert their gas to stars
or if we properly account for their entire baryonic content within $r_e$, then we expect that 
they will satisfy the same scaling relation as all other galaxies. 

\subsection{Revisiting our Simplifications}
\label{sec:revisiting}

Before we proceed to discuss further implications of these results,
we step back to explore how a failure to satisfy 
the assumptions invoked in our simplifications of the virial theorem
would manifest itself in our evaluation of Eq. \ref{eq:6}.
The potential ``failures" fall into three classes. 

First,
we might have introduced errors that are constant  across the galaxy population.
An example of such an error 
would be if we always underestimated the potential energy in our evaluation of the 
virial theorem by a fixed factor.
Such an error would manifest itself as 
a zero point shift of the data relative to the expectation.
Because we do not calculate the specific constant in Eq. \ref{eq:6} from any physical 
argument, this type of 
error will be transparently absorbed into the constant term when we determine it
using independent measurements, as done in \S \ref{sec:validity}.
For almost all of our discussion, this type of error is difficult to detect but irrelevant.

Second, we might have introduced errors that vary
systematically across the galaxy population. An example 
of such an error would be if we underestimated the potential energy by a certain factor 
for low mass systems but overestimated it by a similar factor for high mass systems. Such
errors, to the degree that they correlate with at least one of the structural parameters, will lead
to changes in the coefficients in Eq. \ref{eq:6} or that describe $\Upsilon_e^f$ (Table \ref{tab:ml}),
but would not introduce scatter. This effect is analogous to introducing a ``tilt" in FP analyses.
Identifying this type of error is critical
if one aims to understand the specific nature of the fitted relationships such as that describing 
$\Upsilon_e^f$ or to compare
with simulations. We implicitly tested 
for such effects across galaxy types in \S \ref{sec:simplicity} and across galaxy mass 
in \S \ref{sec:validity}. The coefficients derived from the virial theorem
and a mass-to-light ratio that scales directly with independently-derived mass-to-light ratios
successfully produce a tight scaling relation (Eq. \ref{eq:6} and Fig. \ref{fig:fm}). This result
demonstrates
that there is no effective ``tilt" either with morphological type or across the full range of 
galaxy luminosities.

Third, we might
have introduced errors that are variable and not systematic across the galaxy population. 
An example of such an error
would be if we have ignored a key determining factor of galactic structure, {\sl e.g.}
the number of nearby neighbors. In naive models, close passages affect the luminosity
of the system but do not affect the size or internal kinematics --- leading to potential
outliers in Eq. \ref{eq:6}.
Such effects, to the degree that they do not correlate with the remaining structural parameters, 
will introduce scatter at each point in the $(r_e,I_e,V)$ space. The low scatter in 
both the entire sample (Fig. \ref{fig:fm}) and the subsample with independently-measured
mass-to-light ratios (Fig. \ref{fig:mlfm}) demonstrates that any such errors introduce
little noise.

\subsection{Connecting $\Upsilon_e$ to Physical Parameters}
\label{sec:connecting}

What we have discussed so far are the end-products, or observables $(r_e,I_e,V,\Upsilon_e)$, rather than inputs, or
true physical quantities, that determine the structure of galaxies.
Two natural candidates for the parameters that drive
galactic structure
are the mass and angular momentum of a galaxy. For disks, analytic treatments of
galaxy formation using these two variables
have been relatively successful \citep{fall, dalcanton}. Although these
models require a few key assumptions that may not be entirely accurate, 
such as the conservation of angular momentum during collapse, their success,
in combination with the results presented
here, suggest that simple dynamical models may be able to reproduce
the observable properties of galaxies. However, proceeding from such difficult to 
measure quantities as mass and angular momentum to the observed structure of galaxies in
a single step is likely to prove difficult.

We focus instead on what we have learned so far from our analysis.
Given that the virial theorem plus $\Upsilon_e$ are all one
needs to generate a gross description of galactic structure (\S \ref{sec:beyond} --- \ref{sec:validity}), all of the interesting
astrophysics of galaxy formation --- at least as related to determining the current, gross structure of 
a galaxy --- is encapsulated in $\Upsilon_e$.  What determines differences in $\Upsilon_e$ among galaxies? 

First, galaxies might convert a different fraction, $\eta$, of their baryons to stars. 
Assuming a universal baryon-to-dark matter ratio for halos, this fraction can be measured
using the total mass-to-light ratio, $\Upsilon_{200}$, evaluated at $R_{200}$, the radius
within which the mass density is 200 times the critical density and the system is 
roughly virialized.
Systematic  variations in $\eta$ have already been noted in studies of halo occupation statistics \citep{vdB,yang},  
lensing mass measurements \citep{hoekstra, mandelbaum}, and direct baryon 
measurements \citep{lin,baryons}. 
All of these studies show that $\Upsilon_{200}$, and therefore $\eta$, depend on halo mass.

Second, the stars
might be concentrated to varying degrees relative to the dark matter due to differences
in the assembly history. We quantify 
the stellar concentration, $\xi$, using $\xi \equiv R_{200}/r_e$.
As the stars become more concentrated in the halo, $\xi$ increases and $\Upsilon_e$ 
decreases. 

\begin{figure}
\plotone{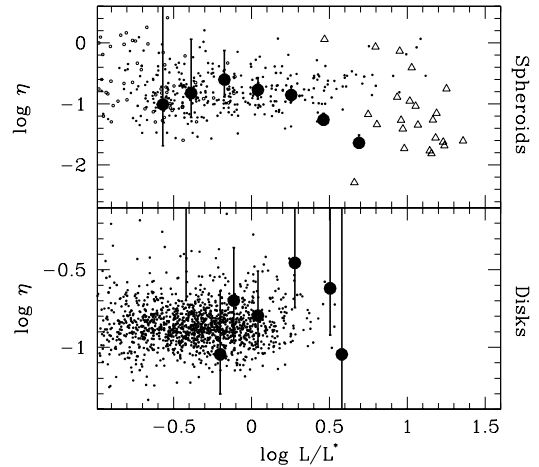}
\caption{Fraction of baryons that are converted to stars, $\eta$, 
as a function of $L/L^*$ for spheroids (upper) and disks (lower). Our data are
the small dots. The values from \cite{mandelbaum} are shown as large filled circles 
with error bars and represent average
values for bins of $L/L^*$. This plot includes the cluster spheroids of \cite{zgz} as open triangles, so that
we extend the range of $L/L^*$ to the higher luminosities (see text) probed by \cite{mandelbaum}.}
\label{fig:mandelbaum}
\end{figure}

We now return to Eq. \ref{eq:2} with the aim of extracting from it expressions for the mass
fraction of baryons that are converted to stars, $\eta$, and the degree
to which the stars are concentrated relative to the dark matter, $\xi$.
We rewrite Eq. \ref{eq:2} as appropriate at the 
virial radius, $R_{200}$,
\begin{equation}
\label{eq:7}
A_{200} V_{200}^2 = {B_{200} GM_{200}\over R_{200}},
\end{equation}
where
$M_{200} \equiv {4\over 3}\pi R_{200}^3\rho_{200}$,
$\rho_{200} \equiv 200\rho_{crit}$, and
$R_{200} \equiv \xi r_e$, 
where $\rho_{crit}$ is the universal critical mass density at the present epoch. To make
further progress, we set $V = V_{200}$. 
This is patently incorrect both because the dark matter potential itself is unlikely to be isothermal
out to $R_{200}$ \citep{nfw}
and any central concentration of the baryons will affect $V$. However, any constant fractional 
differences --- for example, $V = 1.2 V_{200} $ for all galaxies --- will
be absorbed later into our normalization.
What do concern us, but are ignored here, are differences in this velocity scaling that depend on 
the properties of the galaxy \citep{courteau}. This problem might
be correctable in an iterative scheme (i.e., assume a non-isothermal potential, estimate
$\xi$, evaluate the difference between $V$ and $V_{200}$, correct $V_{200}$, and iterate until convergence) or
in a more sophisticated model of galaxy formation \citep{somerville}, but both of these remedies
require some model assumptions and 
lead us from the analytic descriptions that we aim to explore. Work is needed to 
determine the magnitude of the error introduced by our simple treatment.

Continuing, we define $\eta$ through the equation

\begin{equation}
\label{eq:8}
L \equiv {\eta f_B M_{200}\over \Upsilon_{*}},
\end{equation}
where $\Upsilon_{*}$ is the mass-to-light ratio of the stellar population, 
$L$ is the total luminosity, and $f_B$ is the baryon mass fraction.
Algebra enables us to derive equations for $\xi$ and $\eta$:

\begin{equation}
\label{eq:9}
\xi = \Big({{3A_{200} \over 800\pi \rho_{crit} B_{200} G}}\Big)^{1 \over 2} {\rm \ }{V\over r_e}
\end{equation}

and

\begin{equation}
\label{eq:10}
\eta = \Big({{800 \pi B_{200}^3 G^3   \rho_{crit}  \over 3 f_b^2 A_{200}^3}}\Big)^{1 \over 2} {\rm \ } {L \Upsilon_* \over V^3}.
\end{equation}

We express the combination of these two quantities as 
\begin{equation}
\label{eq:11}
\log \xi + \log \eta =  \log {KV\over r_e} + \log {JL \Upsilon_* \over V^3},
\end{equation}
where all constants, as well as the structural factors $A_{200}$ and $B_{200}$,
in Eqs. \ref{eq:9} and \ref{eq:10} are contained in  $K$ and $J$.
$A_{200}$ and $B_{200}$ are not necessarily constant, although their analogs
at $r_e$ are well-behaved (\S \ref{sec:simplicity}). We assume that $A_{200}$ and $B_{200}$ 
are similarly well-behaved
at $R_{200}$.
Rewriting, we get
\begin{equation}
\label{eq:12}
\log \xi + \log \eta =  -2\log V +\log r_e + \log I_e + \log \Upsilon_* + const.
\end{equation}
We know from \S \ref{sec:simplicity} that we can replace the three leading terms on the right hand side
with $-\log \Upsilon_e$ to within a constant, so
\begin{equation}
\label{eq:13}
\log \xi + \log \eta =  -\log \Upsilon_e + \log \Upsilon_* + const.
\end{equation}
Because  $\Upsilon_e = f(V,I_e)$, as defined in
Table 1, $\log \eta
+ \log \xi$ is also a function $V$ and $I_e$.  

To provide a more direct example of the possible use of these equations, we use the
results from \cite{mandelbaum} to evaluate the constant terms (including the assumed 
constant terms $A_{200}$ and $B_{200}$) in Eqs. \ref{eq:9} and \ref{eq:10}. \cite{mandelbaum}
use results from weak lensing to evaluate the fraction of baryons that are turned into
stars as a function of galaxy luminosity. They provide 
empirical values for $\eta$ as a function of galaxy luminosity
and morphology. 
We use their results for $\eta$ for $L^*$ spirals, and assume $\Upsilon_{*} = 1.7$ in the $I$-band
and a universal baryon mass fraction of 0.175 \citep{spergel}, to set the values of
the leading coefficients in 
our Eqs. \ref{eq:9} and \ref{eq:10}: $\xi =  1.4 V r_e^{-1}$ and $\eta = 1.9 \times 10^{-5} (L/L^*)\Upsilon_* V^{-3}$.  
The adopted value of $\Upsilon_*$ 
for an $L^*$ spiral (assuming that it is an
Sab type with $B-V$= 0.8; \citep{fukugita}) is calculated using Table 1 of \cite{bell}.
We then plot
our calculated values of $\eta$ for all of our galaxies (Figure \ref{fig:mandelbaum}), assuming
that $\Upsilon_*$ is 1.7 and 2.5 for disks and spheroids, respectively.  For the
purpose of this comparison, we augment our
galaxy sample with measurements of cluster spheroids (CSph), the brightest cluster galaxy plus
the intracluster stars of groups and clusters \citep{zgz}. 
Our previous work shows that these systems lie on the
FM and, in this context, they allow us to extend the range of $L/L^*$ over
which we can compare to the \cite{mandelbaum} results. We decided against including
the CSphs throughout our current study because our focus is on galaxies, but
CSphs do indeed fall on the 1:1 line in Figure \ref{fig:fm}.

Our values for $\eta$ are in agreement
over the range of luminosities and galaxy types 
presented by \cite{mandelbaum}, except possibly for the  
highest luminosity spheroids: brightest cluster galaxies and CSphs. Our data suggest a qualitatively similar drop in $\eta$ to that
found by \cite{mandelbaum}, although 
at somewhat larger values of $L/L^*$. There are many technical reasons (such as the use of
different bandpasses, subtleties in the
definition of total magnitudes, and complications introduced by intracluster light for these 
most massive spheroids)
that preclude any conclusion about whether there is a true
discrepancy. In general, we
agree quite well both qualitatively and quantitatively with their results. This agreement, in turn,
suggests that $A_{200}$ and $B_{200}$ do not vary strongly as a function of mass or morphological
type.

\begin{figure*}
\plotone{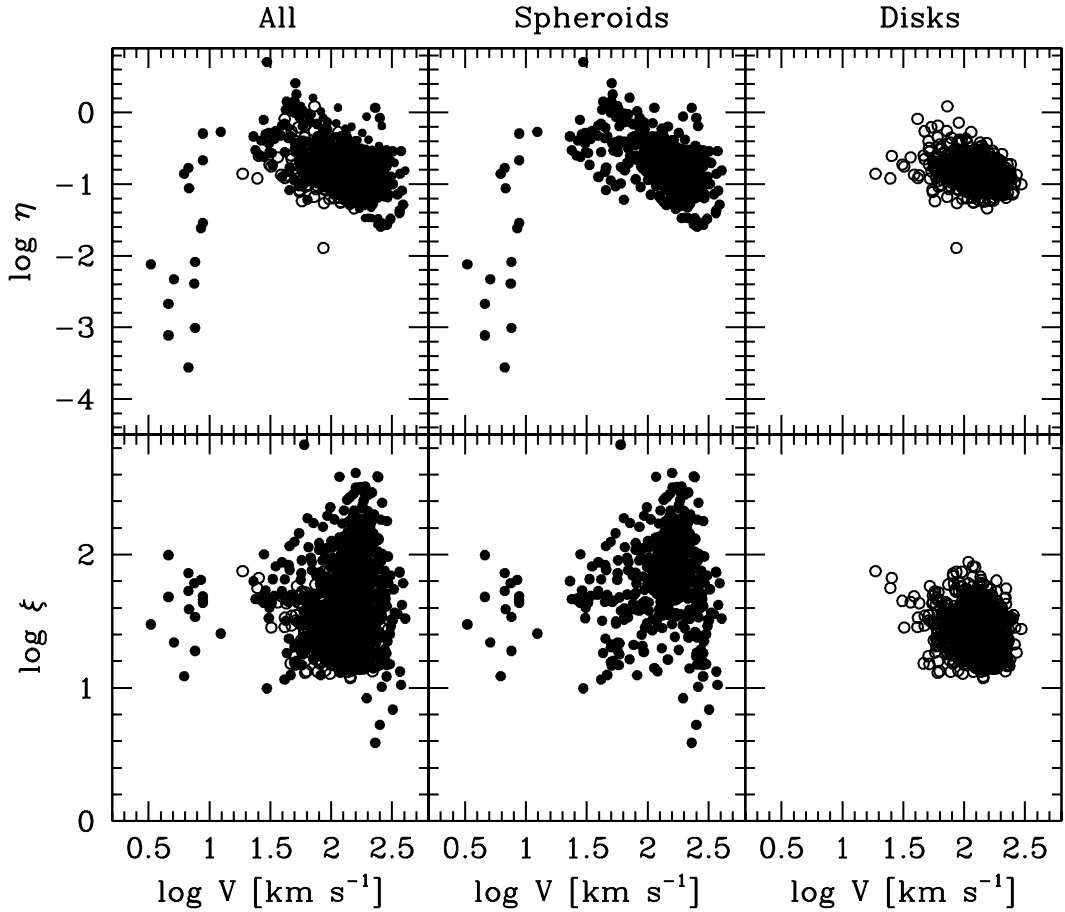}
\caption{Mass fraction of baryons that are converted to stars, $\eta$, (top) 
and stellar concentration, $\xi$, (bottom) as a 
a function of $V$ for our entire sample (spheroids represented with filled circles, disks with open
circles).  Various results are in evidence, including the 
dramatic decrease in $\eta$ for the dSphs, the similarity in $\eta$ among spirals and spheroids
in the regime where they overlap in $V$, the systemic decline in $\eta$ with increasing $V$ for
$\log V > 1.5$, and the generally greater $\xi$ for spheroids relative to disks in that same velocity range.}
\label{fig:structure}
\end{figure*}

We proceed now to calculate $\eta$ and $\xi$ as a function of $V$ for all of our galaxies and show
the results in Figure \ref{fig:structure}. Here, to within the flaws in our
simple derivation and the heterogeneous sample, is a full description of how baryons turn
into stars and distribute themselves in all galaxies ranging from dSphs to BCGs.
There are several striking results. First, the relatively large values of $\Upsilon_e$ for dwarf spheroidals
are primarily driven by $\eta$ rather than by $\xi$, which is
surprisingly constant across the full range of $V$. Second, at a given $V$, ellipticals
and spirals have similar values of $\eta$. Therefore, the difference in their values of $\Upsilon_e$
are due primarily to differences in $\xi$. Third, there is a steady decline in $\eta$ for $\log V > 1.5$ ($V >$ 32 km s$^{-1}$) even for the spheroids, among which large variations in $\Upsilon_*$ are less 
likely.
Systems with comparable $V$ to the Milky Way, $149 < V < 163$ km s$^{-1}$ or alternatively
$210 < v_c < 230$ km s$^{-1}$, have $\eta =  0.14 \pm 0.04$, with the spheroids being
on average 5 times more concentrated than the disks.
All of these results await resolution of two key open questions in the evaluation of $\xi$ and $\eta$:
1) how to treat the difference between $V$ and $V_{200}$ and 2) whether the structural
terms are as well-behaved at $R_{200}$ as they are at $r_e$. Nevertheless, this analysis
illustrates how we might construct a bridge between the empirical relations based
on observables and more theoretical ones based on fundamental physical parameters.
Thus, we achieve the third goal listed in \S \ref{sec:intro}.

\section{Summary}
\label{sec:summary}

We have shown that all classes of galaxies, ranging in mass from dwarf spheroidals to brightest 
cluster galaxies, and in type from spheroids to disks, fall on a two dimensional
surface in the observable space  $(r_e,I_e,V)$, where $ V^2 \equiv {1\over 2} v_c^2 + 
\sigma^2$, over three orders of magnitude in $r_e$ with $<$ 25\%
scatter. The scatter about that surface is comparable to that observed in Fundamental
Plane and Tully-Fisher studies in which the range of galaxy types and
luminosities is much more limited. The TF and FP relationships are subsets of
the manifold presented here. This finding alone demonstrates that the structure of all galaxies
can be described with a highly limited set of parameters. The observational ones, $(r_e,I_e,V)$,
may not be optimal, even though they do a remarkably good job. The small scatter about
the mean relation implies that a host of potential physical phenomena such as 
environmental effects, star formation history, nuclear activity, accretion history, and feedback
are either (1) relatively unimportant in determining the structure of galaxies, (2) 
move galaxies along this well-defined relationship, or (3) balance each other so as to
define the mean relation as the locus of galactic structure equilibria. 

We developed a simple analytic treatment in which we 
asserted the existence of a fundamental manifold of galaxies. By requiring the
simple virial theorem derivation to result in a two dimensional manifold in observed space, 
we specify the behavior of the mass-to-light ratio within $r_e$, $\Upsilon_e$. We then tested this assertion by comparing our inferred 
values of $\Upsilon_e$ to those derived independently from much more sophisticated 
modeling for both normal ellipticals and dSphs. The agreement is quantitatively excellent, with
less than 15\% scatter in mass-to-light ratios for those galaxies that satisfy our dynamical
criteria. This result demonstrates that the principal additional ingredient necessary in 
proceeding from the virial theorem to a description of galactic structure is knowledge of
the mass-to-light ratio within $r_e$. Additional factors, which could have been important,
such as internal kinematic anisotropy or variations in the radial profile of the gravitational
potential from one galaxy to another, must play a role at less than the 25\% level.
The observed manifold is described by
\begin{equation}
\label{eq:14}
\log r_e -  \log V^2 + \log I_e + \log \Upsilon_e + 0.8 = 0,
\end{equation}
where we also provide a fitting function for $\log \Upsilon_e$  in terms of $V$ and $I_e$.
The equations presented here are numerically appropriate for $r_e$, $V$, $I_e$, and $\Upsilon$
in units of kpc, km s$^{-1}$, $L_\odot/$pc$^{-2}$, and solar units, respectively, and based primarily
on $I$-band observations. 

We then discuss what the inferred behavior of $\Upsilon_e$ may mean for
the physical characteristics of the galaxies. In particular, we speculate that
the two principal determinants of $\Upsilon_e$ are 
the mass fraction of baryons that are turned into stars, $\eta$,
and the degree to which the stars are spatially concentrated relative to the dark
matter, $\xi \equiv R_{200}/r_e$. We derive equations for both quantities in 
terms of unknown structural parameters and the observables. We relate
 the two quantities
using the expression that we derived for $\Upsilon_e$. 
Finally, we use independent measures of $\eta$ \citep{mandelbaum} to solve for the unknown structural 
terms for one set of galaxies and then compare the behavior of $\eta$ across
other luminosities and galaxy types as determined both from our analysis and
that independent weak lensing study. This comparison
leads to simple expressions for $\eta$ and $\xi$,
\begin{equation}
\eta = 1.9 \times 10^{-5} {(L/L^*) \Upsilon_* \over V^3}
\label{eq:15}
\end{equation}
and
\begin{equation}
\label{eq:16}
\xi = 1.4 {V \over r_e}.
\end{equation}
We 
are then able to extend the measurements of $\eta$ and $\xi$ to the full range of galaxies.
As rough guides, we find
that,  for most galaxies, $0.04< \eta < 0.6$ and $10 < \xi < 200$, although these can 
be evaluated on a galaxy-by-galaxy basis.  
Systems with comparable $V$ to the Milky Way, $149 < V < 163$ km s$^{-1}$ or alternatively
$210 < v_c < 230$ km s$^{-1}$, have $\eta =  0.14 \pm 0.04$, with the spheroids being
on average 5 times more concentrated than the disks.
Overall, we reach a set of general conclusions. 
First, the relatively large values of $\Upsilon_e$ for dwarf spheroidals
are primarily driven by $\eta$ rather than by $\xi$, which is
surprisingly constant across the full range of $V$. Second, at a given $V$, ellipticals
and spirals have similar values of $\eta$ ($<$ 10\% difference for spheroids and disks with
$149 < V < 163$ km s$^{-1}$). Therefore, the difference in their values of $\Upsilon_e$
is due primarily  to differences in $\xi$. Third, there is a steady decline in $\eta$ for $\log V > 1.5$ 
($V >$ 32 km s$^{-1}$).

The data used here fall short of the ideal sample from which to properly derive 
the quantitative values that mathematically describe the manifold, primarily due to the
heterogeneous nature of the amalgamated sample. Nevertheless, the sample does
demonstrate that the range of galaxy structure is dominated by only two parameters. The lower
scatter obtained either for a single sample (0.06 dex for the \cite{springob} disk sample) or for
independently derived $\Upsilon_e$'s (0.04 dex when using both the \cite{cappellari}
data for ellipticals and the \cite{walker} data for dSphs) suggest that a homogeneous
sample might show that the myriad of 
possible influences on galactic structure (environment, 
accretion history, AGN activity, star formation history, and others) contribute at most a 
$\sim$ 10\% scatter to the scaling
relationship presented in Eq. \ref{eq:14}.

The existence of a highly constrained surface on which galaxies lie does not 
eliminate the need for additional physics. In particular, as we have hinted throughout, 
many physical effects might move galaxies along the surface or perhaps counter-balance to move galaxies back to the equilibrium surface described by the manifold. 
As such, future galaxy
models may be more constrained by the distribution of galaxies on the surface rather
than perpendicular to it. Our heterogeneous sample is ill-suited to say much about the 
distribution of sources on the surface. Much work still remains.

We close by returning to the analogy of stellar structure. It is evident that we are
still far from a physical theory of galactic structure, but that we have progressed in 
several key aspects. First, we have now demonstrated that the entire family of
galaxies can be described by sets of two parameters (e.g., $V$ and $I_e$ or $\eta$ and $\xi$). 
This finding 
motivates the search for relatively simple expressions of galactic structure that
are connected to a small set of physical parameters, such as mass and angular 
momentum. Second, we have identified the principal characteristic that remains
to be explained, namely $\Upsilon_e$. The virial theorem plus
an understanding of $\Upsilon_e$ are all that are necessary to predict the size, internal kinematics, or
luminosity of a galaxy, when given the other two. This in turn places
the focus on understanding what determines the fraction of baryons that are turned
into stars and how those stars are packed within the dark halo. If those quantities can then be 
connected to more fundamental parameters, such as mass and angular momentum, then
one could proceed from the physical parameters directly to the observables. 
At that point, we will have indeed produced equations of galactic structure.

\begin{acknowledgments}

DZ acknowledges financial support for this work from a Guggenheim fellowship,
NASA LTSA award NNG05GE82G, and NSF grant AST-0307482.
AIZ acknowledges financial support from NASA  awards LTSA NAG5-11108 and
ADP NNG05GC94G, and 
from NSF grant AST-0206084. DZ and AZ also want to thank the NYU Physics
department and the Center for Cosmology and Particle Physics for their generous
support and hospitality during their sabbatical there.

\end{acknowledgments}
                         
\bibliographystyle{apj}
\bibliography{ms}

\begin{thebibliography}{40}
\expandafter\ifx\csname natexlab\endcsname\relax\def\natexlab#1{#1}\fi

\bibitem[{{Bell} \& {de Jong}(2001)}]{bell}
{Bell}, E.~F., \& {de Jong}, R.~S. 2001, \apj, 550, 212

\bibitem[{{Bernardi} {et~al.}(2003){Bernardi}, {Sheth}, {Annis}, {Burles},
  {Eisenstein}, {Finkbeiner}, {Hogg}, {Lupton}, {Schlegel}, {SubbaRao},
  {Bahcall}, {Blakeslee}, {Brinkmann}, {Castander}, {Connolly}, {Csabai},
  {Doi}, {Fukugita}, {Frieman}, {Heckman}, {Hennessy}, {Ivezi{\'c}}, {Knapp},
  {Lamb}, {McKay}, {Munn}, {Nichol}, {Okamura}, {Schneider}, {Thakar}, \&
  {York}}]{bernardi}
{Bernardi}, M., {Sheth}, R.~K., {Annis}, J., {Burles}, S., {Eisenstein}, D.~J.,
  {Finkbeiner}, D.~P., {Hogg}, D.~W., {Lupton}, R.~H., {Schlegel}, D.~J.,
  {SubbaRao}, M., {Bahcall}, N.~A., {Blakeslee}, J.~P., {Brinkmann}, J.,
  {Castander}, F.~J., {Connolly}, A.~J., {Csabai}, I., {Doi}, M., {Fukugita},
  M., {Frieman}, J., {Heckman}, T., {Hennessy}, G.~S., {Ivezi{\'c}}, {\v Z}.,
  {Knapp}, G.~R., {Lamb}, D.~Q., {McKay}, T., {Munn}, J.~A., {Nichol}, R.,
  {Okamura}, S., {Schneider}, D.~P., {Thakar}, A.~R., \& {York}, D.~G. 2003,
  \aj, 125, 1866

\bibitem[{{Bothun}(1984)}]{bothun}
{Bothun}, G.~D. 1984, \apj, 277, 532

\bibitem[{{Burstein} {et~al.}(1997){Burstein}, {Bender}, {Faber}, \&
  {Nolthenius}}]{burstein}
{Burstein}, D., {Bender}, R., {Faber}, S., \& {Nolthenius}, R. 1997, \aj, 114,
  1365

\bibitem[{{Cappellari} {et~al.}(2006){Cappellari}, {Bacon}, {Bureau}, {Damen},
  {Davies}, {de Zeeuw}, {Emsellem}, {Falc{\'o}n-Barroso}, {Krajnovi{\'c}},
  {Kuntschner}, {McDermid}, {Peletier}, {Sarzi}, {van den Bosch}, \& {van de
  Ven}}]{cappellari}
{Cappellari}, M., {Bacon}, R., {Bureau}, M., {Damen}, M.~C., {Davies}, R.~L.,
  {de Zeeuw}, P.~T., {Emsellem}, E., {Falc{\'o}n-Barroso}, J., {Krajnovi{\'c}},
  D., {Kuntschner}, H., {McDermid}, R.~M., {Peletier}, R.~F., {Sarzi}, M., {van
  den Bosch}, R.~C.~E., \& {van de Ven}, G. 2006, \mnras, 366, 1126

\bibitem[{{Cappellari} {et~al.}(2007){Cappellari}, {Emsellem}, {Bacon},
  {Bureau}, {Davies}, {de Zeeuw}, {Falc{\'o}n-Barroso}, {Krajnovi{\'c}},
  {Kuntschner}, {McDermid}, {Peletier}, {Sarzi}, {van den Bosch}, \& {van de
  Ven}}]{cappellari2007}
{Cappellari}, M., {Emsellem}, E., {Bacon}, R., {Bureau}, M., {Davies}, R.~L.,
  {de Zeeuw}, P.~T., {Falc{\'o}n-Barroso}, J., {Krajnovi{\'c}}, D.,
  {Kuntschner}, H., {McDermid}, R.~M., {Peletier}, R.~F., {Sarzi}, M., {van den
  Bosch}, R.~C.~E., \& {van de Ven}, G. 2007, \mnras, 379, 418

\bibitem[{{Courteau} {et~al.}(2007){Courteau}, {McDonald}, {Widrow}, \&
  {Holtzman}}]{courteau}
{Courteau}, S., {McDonald}, M., {Widrow}, L.~M., \& {Holtzman}, J. 2007, \apjl,
  655, L21

\bibitem[{{Dalcanton} {et~al.}(1997){Dalcanton}, {Spergel}, \&
  {Summers}}]{dalcanton}
{Dalcanton}, J.~J., {Spergel}, D.~N., \& {Summers}, F.~J. 1997, \apj, 482, 659

\bibitem[{{Djorgovski} \& {Davis}(1987)}]{dd87}
{Djorgovski}, S., \& {Davis}, M. 1987, \apj, 313, 59

\bibitem[{{Dressler} {et~al.}(1987){Dressler}, {Lynden-Bell}, {Burstein},
  {Davies}, {Faber}, {Terlevich}, \& {Wegner}}]{d87}
{Dressler}, A., {Lynden-Bell}, D., {Burstein}, D., {Davies}, R.~L., {Faber},
  S.~M., {Terlevich}, R., \& {Wegner}, G. 1987, \apj, 313, 42

\bibitem[{{Faber} \& {Gallagher}(1979)}]{faber}
{Faber}, S.~M., \& {Gallagher}, J.~S. 1979, \araa, 17, 135

\bibitem[{{Faber} \& {Jackson}(1976)}]{fj}
{Faber}, S.~M., \& {Jackson}, R.~E. 1976, \apj, 204, 668

\bibitem[{{Fall} \& {Efstathiou}(1980)}]{fall}
{Fall}, S.~M., \& {Efstathiou}, G. 1980, \mnras, 193, 189

\bibitem[{{Fukugita} {et~al.}(1995){Fukugita}, {Shimasaku}, \&
  {Ichikawa}}]{fukugita}
{Fukugita}, M., {Shimasaku}, K., \& {Ichikawa}, T. 1995, \pasp, 107, 945

\bibitem[{{Gavazzi} {et~al.}(2007){Gavazzi}, {Treu}, D., {Koopmans}, {Bolton},
  {Burles}, {Massey}, \& {Moustakas}}]{gavazzi}
{Gavazzi}, R., {Treu}, T., D., R.~J., {Koopmans}, L.~V.~E., {Bolton}, A.~S.,
  {Burles}, S., {Massey}, R., \& {Moustakas}, L.~A. 2007, \apj, in press

\bibitem[{{Geha} {et~al.}(2006){Geha}, {Blanton}, {Masjedi}, \&
  {West}}]{geha2006}
{Geha}, M., {Blanton}, M.~R., {Masjedi}, M., \& {West}, A.~A. 2006, \apj, 653,
  240

\bibitem[{{Gonzalez} {et~al.}(2007){Gonzalez}, {Zaritsky}, \&
  {Zabludoff}}]{baryons}
{Gonzalez}, A.~H., {Zaritsky}, D., \& {Zabludoff}, A.~I. 2007, \apj, 666, 147

\bibitem[{{Hoekstra} {et~al.}(2005){Hoekstra}, {Hsieh}, {Yee}, {Lin}, \&
  {Gladders}}]{hoekstra}
{Hoekstra}, H., {Hsieh}, B.~C., {Yee}, H.~K.~C., {Lin}, H., \& {Gladders},
  M.~D. 2005, \apj, 635, 73

\bibitem[{{J\o rgensen} {et~al.}(1996){J\o rgensen}, {Franx}, \&
  {Kjaergaard}}]{jorg}
{J\o rgensen}, I., {Franx}, M., \& {Kjaergaard}, P. 1996, \mnras, 280, 167

\bibitem[{{Kannappan}(2004)}]{kannappan}
{Kannappan}, S.~J. 2004, \apjl, 611, L89

\bibitem[{{Kassin} {et~al.}(2007){Kassin}, {Weiner}, {Faber}, {Koo}, {Lotz},
  {Diemand}, {Harker}, {Bundy}, {Metevier}, {Phillips}, {Cooper}, {Croton},
  {Konidaris}, {Noeske}, \& {Willmer}}]{kassin}
{Kassin}, S.~A., {Weiner}, B.~J., {Faber}, S.~M., {Koo}, D.~C., {Lotz}, J.~M.,
  {Diemand}, J., {Harker}, J.~J., {Bundy}, K., {Metevier}, A.~J., {Phillips},
  A.~C., {Cooper}, M.~C., {Croton}, D.~J., {Konidaris}, N., {Noeske}, K.~G., \&
  {Willmer}, C.~N.~A. 2007, \apjl, 660, L35

\bibitem[{{Lin} {et~al.}(2004){Lin}, {Mohr}, \& {Stanford}}]{lin}
{Lin}, Y.-T., {Mohr}, J.~J., \& {Stanford}, S.~A. 2004, \apj, 610, 745

\bibitem[{{Mandelbaum} {et~al.}(2006){Mandelbaum}, {Seljak}, {Kauffmann},
  {Hirata}, \& {Brinkmann}}]{mandelbaum}
{Mandelbaum}, R., {Seljak}, U., {Kauffmann}, G., {Hirata}, C.~M., \&
  {Brinkmann}, J. 2006, \mnras, 368, 715

\bibitem[{{McGaugh}(2005)}]{mcgaugh05}
{McGaugh}, S.~S. 2005, \apj, 632, 859

\bibitem[{{McGaugh} {et~al.}(2000){McGaugh}, {Schombert}, {Bothun}, \& {de
  Blok}}]{mcgaugh00}
{McGaugh}, S.~S., {Schombert}, J.~M., {Bothun}, G.~D., \& {de Blok}, W.~J.~G.
  2000, \apjl, 533, L99

\bibitem[{{Navarro} {et~al.}(1997){Navarro}, {Frenk}, \& {White}}]{nfw}
{Navarro}, J.~F., {Frenk}, C.~S., \& {White}, S.~D.~M. 1997, \apj, 490, 493

\bibitem[{{Pizagno} {et~al.}(2007){Pizagno}, {Prada}, {Weinberg}, {Rix},
  {Pogge}, {Grebel}, {Harbeck}, {Blanton}, {Brinkmann}, \& {Gunn}}]{pizagno}
{Pizagno}, J., {Prada}, F., {Weinberg}, D.~H., {Rix}, H.-W., {Pogge}, R.~W.,
  {Grebel}, E.~K., {Harbeck}, D., {Blanton}, M., {Brinkmann}, J., \& {Gunn},
  J.~E. 2007, \aj, 134, 945

\bibitem[{{Read} \& {Trentham}(2005)}]{read}
{Read}, J.~I., \& {Trentham}, N. 2005, Royal Society of London Philosophical
  Transactions Series A, 363, 2693

\bibitem[{{Roberts}(1969)}]{roberts}
{Roberts}, M.~S. 1969, \aj, 74, 859

\bibitem[{{Simon} \& {Geha}(2007)}]{simon}
{Simon}, J.~D., \& {Geha}, M. 2007, ArXiv e-prints/0706.0516

\bibitem[{{Somerville} {et~al.}(2007){Somerville}, {Barden}, {Rix}, {Bell},
  {Beckwith}, {Borch}, {Caldwell}, {Haeussler}, {Heymans}, {Jahnke}, {Joghee},
  {McIntosh}, {Meisenheimer}, {Peng}, {Sanchez}, {Wisotzki}, \&
  {Wolf}}]{somerville}
{Somerville}, R., {Barden}, M., {Rix}, H.-W., {Bell}, E.~F., {Beckwith}, S.,
  {Borch}, A., {Caldwell}, J., {Haeussler}, B., {Heymans}, C., {Jahnke}, K.,
  {Joghee}, S., {McIntosh}, D., {Meisenheimer}, K., {Peng}, C., {Sanchez}, S.,
  {Wisotzki}, L., \& {Wolf}, C. 2007, ArXiv e-prints/0612428

\bibitem[{{Spergel} {et~al.}(2007){Spergel}, {Bean}, {Dor{\'e}}, {Nolta},
  {Bennett}, {Dunkley}, {Hinshaw}, {Jarosik}, {Komatsu}, {Page}, {Peiris},
  {Verde}, {Halpern}, {Hill}, {Kogut}, {Limon}, {Meyer}, {Odegard}, {Tucker},
  {Weiland}, {Wollack}, \& {Wright}}]{spergel}
{Spergel}, D.~N., {Bean}, R., {Dor{\'e}}, O., {Nolta}, M.~R., {Bennett}, C.~L.,
  {Dunkley}, J., {Hinshaw}, G., {Jarosik}, N., {Komatsu}, E., {Page}, L.,
  {Peiris}, H.~V., {Verde}, L., {Halpern}, M., {Hill}, R.~S., {Kogut}, A.,
  {Limon}, M., {Meyer}, S.~S., {Odegard}, N., {Tucker}, G.~S., {Weiland},
  J.~L., {Wollack}, E., \& {Wright}, E.~L. 2007, \apjs, 170, 377

\bibitem[{{Springob} {et~al.}(2007){Springob}, {Masters}, {Haynes},
  {Giovanelli}, \& {Marinoni}}]{springob}
{Springob}, C.~M., {Masters}, K.~L., {Haynes}, M.~P., {Giovanelli}, R., \&
  {Marinoni}, C. 2007, \apj

\bibitem[{{Tully} \& {Fisher}(1977)}]{tf}
{Tully}, R.~B., \& {Fisher}, J.~R. 1977, \aap, 54, 661

\bibitem[{{van den Bosch} {et~al.}(2003){van den Bosch}, {Yang}, \& {Mo}}]{vdB}
{van den Bosch}, F.~C., {Yang}, X., \& {Mo}, H.~J. 2003, \mnras, 340, 771

\bibitem[{{Walker} {et~al.}(2007){Walker}, {Mateor}, {Olszewski}, {Gnedin},
  {Wang}, {Sen}, \& {Woodroofe}}]{walker}
{Walker}, M.~G., {Mateor}, M., {Olszewski}, E.~W., {Gnedin}, O.~Y., {Wang}, X.,
  {Sen}, B., \& {Woodroofe}, M. 2007, \apjl

\bibitem[{{Willick}(1994)}]{willick}
{Willick}, J.~A. 1994, \apjs, 92, 1

\bibitem[{{Yang} {et~al.}(2005){Yang}, {Mo}, {Jing}, \& {van den Bosch}}]{yang}
{Yang}, X., {Mo}, H.~J., {Jing}, Y.~P., \& {van den Bosch}, F.~C. 2005, \mnras,
  358, 217

\bibitem[{{Zaritsky} {et~al.}(2006{\natexlab{a}}){Zaritsky}, {Gonzalez}, \&
  {Zabludoff}}]{zgz}
{Zaritsky}, D., {Gonzalez}, A.~H., \& {Zabludoff}, A.~I. 2006{\natexlab{a}},
  \apj, 638, 725

\bibitem[{{Zaritsky} {et~al.}(2006{\natexlab{b}}){Zaritsky}, {Gonzalez}, \&
  {Zabludoff}}]{zgzb}
---. 2006{\natexlab{b}}, \apjl, 642, L37

\end{thebibliography}

\end{document}